\documentclass[conference]{IEEEtran}
\IEEEoverridecommandlockouts

\usepackage{multirow}
\usepackage{graphicx}

\usepackage{amsmath,graphicx,bbding}
\usepackage{indentfirst}
\usepackage{booktabs}
\usepackage{float}
\usepackage[table,xcdraw]{xcolor}

\usepackage{cite}
\usepackage[numbers,sort&compress]{natbib}
\usepackage{amsmath,amssymb,amsfonts}
\usepackage{algorithmic}
\usepackage{graphicx}
\usepackage{textcomp}
\usepackage{xcolor}
\def\BibTeX{{\rm B\kern-.05em{\sc i\kern-.025em b}\kern-.08em
    T\kern-.1667em\lower.7ex\hbox{E}\kern-.125emX}}
\begin{document}

\title{NLCUnet: Single-Image Super-Resolution Network with Hairline Details}

\author{\IEEEauthorblockN{1\textsuperscript{st} Jiancong Feng}
\IEEEauthorblockA{\textit{Guangzhou University} \\
Guangzhou, China \\
fengjiancong@e.gzhu.edu.cn}
\and
\IEEEauthorblockN{2\textsuperscript{nd} Yuan-Gen Wang$^{\ast}$\thanks{$^{\ast}$denotes the corresponding author. This work was supported by the National Natural Science Foundation of China (Grant Nos. 62272116 and 61872099). The authors acknowledge the Network Center of Guangzhou University for providing HPC computing resources. The code is available at https://github.com/GZHU-DVL/NLCUnet.}}
\IEEEauthorblockA{\textit{Guangzhou University} \\
Guangzhou, China \\
wangyg@gzhu.edu.cn}
\and
\IEEEauthorblockN{3\textsuperscript{rd} Fengchuang Xing}
\IEEEauthorblockA{\textit{Guangzhou University} \\
Guangzhou, China \\
xfchuang@e.gzhu.edu.cn}
}

\maketitle

\footnote{This paper has been accepted for presentation in IEEE ICME 2023.}

\begin{abstract}
Pursuing the precise details of super-resolution images is challenging for single-image super-resolution tasks. This paper presents a single-image super-resolution network with hairline details (termed NLCUnet), including three core designs. Specifically, a non-local attention mechanism is first introduced to restore local pieces by learning from the whole image region. Then, we find that the blur kernel trained by the existing work is unnecessary. Based on this finding, we create a new network architecture by integrating depth-wise convolution with channel attention without the blur kernel estimation, resulting in a performance improvement instead. Finally, to make the cropped region contain as much semantic information as possible, we propose a random 64$\times$64 crop inside the central 512$\times$512 crop instead of a direct random crop inside the whole image of 2K size. Numerous experiments conducted on the benchmark DF2K dataset demonstrate that our NLCUnet performs better than the state-of-the-art in terms of the PSNR and SSIM metrics and yields visually favorable hairline details.
\end{abstract}

\section{Introduction}
Deep convolution neural networks (CNNs) have recently achieved outstanding results in single-image super-resolution (SISR). Existing methods \cite{1,2,3} usually presume that the blur kernel is already defined as the bicubic interpolation kernel. This allows for the manual synthesis of a huge number of training samples, which can be utilized to train an effective CNN. However, the blur kernels in the real world are quite different from the bicubic interpolation kernels, yielding a domain gap between them. When these methods designed by bicubic blur kernels are applied to real scenarios, the domain gap will result in a significant performance drop. Therefore, super-resolution (SR) in the setting of unknown blur kernels, also known as blind SR, attracts great attention in literature.

\indent
Most of blind SR methods introduce an unknown variable called the blur kernel $k$, making the optimization problem challenging. In order to simplify the problem, earlier approaches \cite{4,5,6} often divide the optimization procedure into two consecutive steps: (1) estimating the blur kernel from the low-resolution (LR) image, and (2) restoring the SR image based on the estimated kernel. Two models trained independently may not be fully compatible in the above two-step procedure. On one hand, a minor inaccurate estimation  might have a significant impact on the performance of the subsequent step \cite{7}. On the other hand, the first step only makes use of a small amount of information from the LR image, making it challenging to predict a highly accurate blur kernel. Because of this, even these two models can perform well on their own training step, the resulting performance may not be good when they are combined together. In order to make the two models more compatible and more likely to produce superior performance than the above two-step methods, the authors \cite{8} designed an alternating optimization method between the blur kernel estimation and the SR image restoration. 

\indent
Even though the method \cite{8} has achieved a progress, we reveal that the blur kernel estimation is actually unnecessary. Based on this, we construct a new network by integrating depth-wise convolution with channel attention without the blur kernel estimation, termed NLCUnet in this paper. Moreover, we introduce a non-local attention mechanism to capture long-range dependency of the whole image. In order to make the cropped area include more semantic information, we process the DF2K dataset (DIV2K \cite{9} + Flickr2K \cite{10}) using a center crop (512$\times$512 size) followed by a random crop (64$\times$64 size). Various experiments show that our network not only requires fewer training epochs and but achieves better performance compared with the existing methods.


\begin{figure*}\tiny
\centering
\includegraphics[scale=0.45]{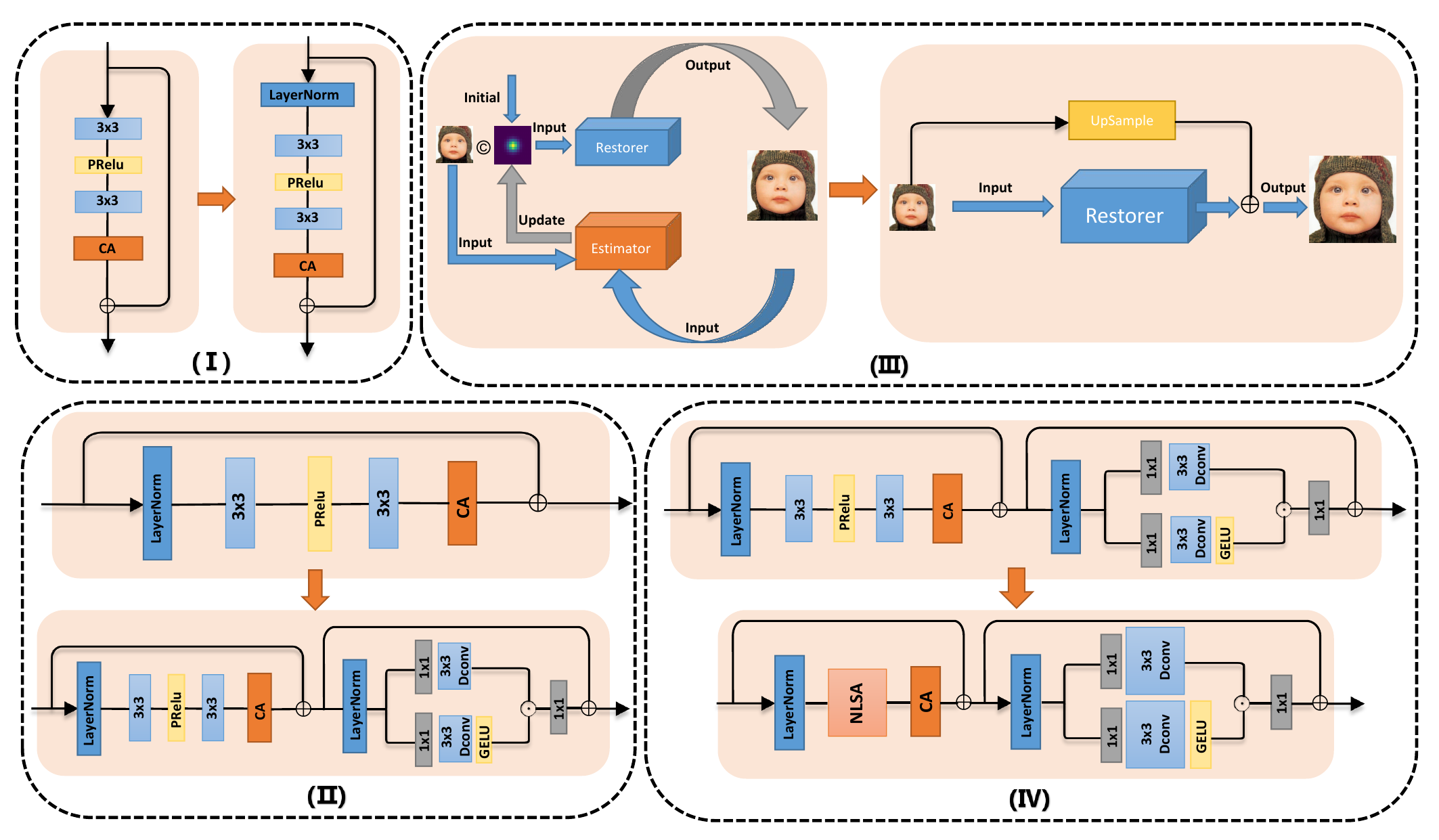}
\vspace{-0.3cm}
\caption{Illustration of our modified network architecture  from DANv1 \cite{8}}
\label{fig1:env}
\end{figure*}

\section{Proposed Method}
\label{sec:method}

We describe the image degradation process as follows:
\begin{equation}
    I_{LR} = (I_{HR} \otimes K) \downarrow _{bic} + N,
\end{equation}
\noindent where $I_{LR}$ denotes the LR image, $I_{HR}$ represents the high-resolution (HR) image, $\otimes$ is the convolution operation, \emph{K} stands for the blur kernel, $\downarrow _{bic}$ indicates the bicubic down-sampling, and \emph{N} denotes additive white Gaussian noise. According to the experiment, we find that these methods \cite{4,5,6,7,8} can achieve similar or better performance without initializing the blur kernel. Therefore, we simplify the process of SR reconstruction to the following equation:
\begin{equation}
    I_{SR} = F(I_{LR}),
\end{equation}
\noindent where $I_{SR}$ denotes the SR image, $F(\cdot)$ is the proposed method in this paper. By training, $F$ can learn the mapping relationship from LR to HR directly without additional blur kernel estimation.

\begin{figure*}\tiny
\centering
\includegraphics[scale=0.5]{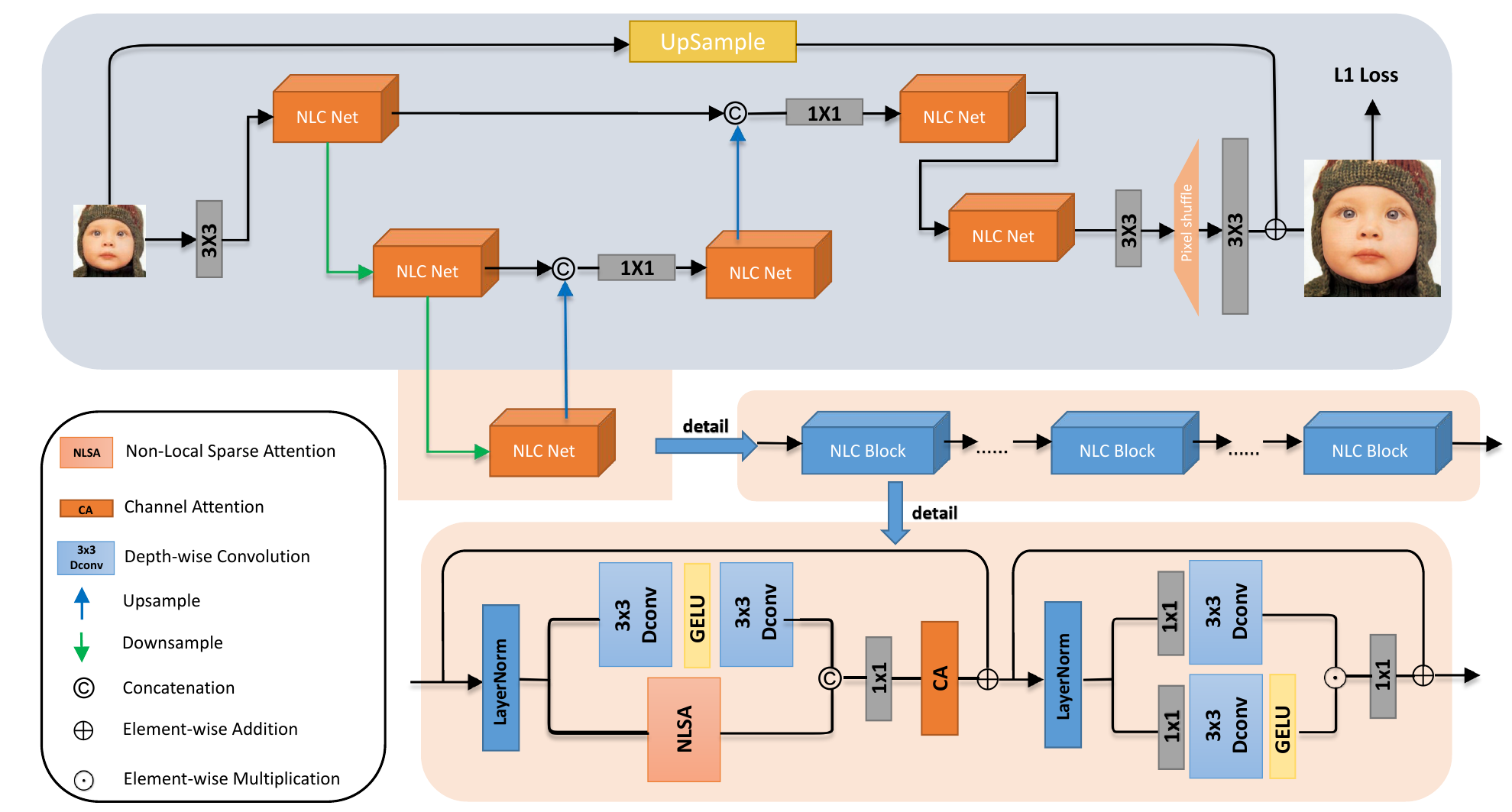}
\vspace{-0.3cm}
\caption{The overview of the proposed NLCUnet.}
\label{fig2:env}
\end{figure*}

The study \cite{11} shows that the CALayer \cite{8} easily causes the gradient explosion during the training. To solve this problem, we propose to perform the LayerNorm operation on the feature map on the channel, as illustrated in (I) of \textbf{Fig. 1}. On the basis of (I), we add a Gated Dconv Feed-Forward Network (GDFN) block \cite{12}, which is equivalent to a nonlinear activation function, as illustrated in (II) of \textbf{Fig. 1}. Moreover, we remove the Estimator of the blur kernel in \cite{8} and add a bicubic interpolation upsampling in the Restorer, as shown in (III) of \textbf{Fig. 1}.

\indent
According to \cite{13}, increasing the depth of the neural network by stacking 3$\times$3 convolutions enlarges only the theoretical receptive fields, not the actual ones. To solve this problem, as shown in (IV) of \textbf{Fig. 1}, our network employs non-local attention (NLA) \cite{14} module to capture long-range dependencies instead of 3$\times$3 convolution. Although the NLA is able to capture long-range dependencies, it ignores the communication across channels. For this aim, we add channel attention (CA) \cite{15} after the NLA to obtain the information interaction between channels. Although adding the NLA blocks can improve performance, the computational complexity of the model increases substantially. Therefore, we replace the NLA blocks with non-local sparse attention (NLSA) \cite{16} blocks to reduce the complexity of the model.

\indent
To preserve the inductive bias property of convolution, we propose a two-branch structure. One branch goes through an NLSA block and the other branch goes through two layers of 3$\times$3 depth-wise convolution. Then we concatenate these two branches on the channel. Since neither of the NLSA block and the depth-wise convolution takes into account the information interaction between the channels, we propose to fuse the two branches with a 1$\times$1 convolution and then use CA for the information interaction between the channels. To reduce the complexity of the network caused by depth increase and to allow the model to learn feature map at different scales, the Unet \cite{17} structure is introduced to our model. Hence, in this paper we name the proposed network NLCUnet (\underline{N}on-local \& \underline{L}ocal \& \underline{C}hannel \underline{Unet}), as shown in \textbf{Fig. 2}.

\indent
As we know GAN can generate good visual quality of the images. Hence, we take our NLCUnet as a generator and the UNet discriminator with spectral normalization \cite{18} as a discriminator, termed NLCUnetGAN in this paper. Our training process is divided into two stages. First, we train a PSNR-oriented model with the $L_1$ loss. We then use the trained PSNR-oriented model as an initialization of the generator, and train the GAN-oriented model with a combination of the $L_1$ loss, perceptual loss, and adversarial loss. Finally, the total loss of our network is represented as follows: 
\begin{equation}
    L_{total} = L_1 + L_{Perc} + 0.1 \times L_{adv}
\end{equation}
\noindent
where $L_1$, $L_{Perc}$, and $L_{adv}$ stands for $L_1$ loss,  perceptual loss, and adversarial loss, respectively. By our experiment, their weights are respectively set to 1, 1, and 0.1. We use the conv1,..., conv5 feature maps before activation of pre-trained VGG19 (weights 0.1, 0.1, 1, 1, 1) as the perceptual loss. 

Besides, we have observed that for most images in the DF2K dataset, the central area often contains more semantic information and is more easily focused by people. Therefore, we propose to first crop the 512$\times$512 block at the center of the image and then perform a random 64$\times$64 crop inside the cropped 512$\times$512 block. In doing so, our network can reduce training epochs and further improve performance.

\section{Experimental Result}
\label{sec:experiment}

\subsection{Experimental configuration}

In this paper, we follow the experimental configuration in \cite{8}, where the DF2K dataset contains 3,450 HR images and is used as the training set. Configuration 1 and Configuration 2 are listed as follows. Configuration 1 is the option in \cite{8}, that exclusively considers situations with isotropic Gaussian blur kernels. Configuration 2 is the option in \cite{19}, which focuses on situations where the blur kernels are irregular and general.

\begin{table*}[]\tiny
\centering
\caption{Quantitative comparison with three leading blind SR methods with Configuration 1. The top-1 and top-2 results are highlighted in red and blue, respectively.}
\label{tab:my-table}
\resizebox{\textwidth}{!}{%
\begin{tabular}{ccccccc}
\hline
{Method} &
  {Scale} &
  {\begin{tabular}[c]{@{}c@{}}Set5\\ PSNR\qquad SSIM\end{tabular}} &
  {\begin{tabular}[c]{@{}c@{}}Set14\\ PSNR\qquad SSIM\end{tabular}} &
  {\begin{tabular}[c]{@{}c@{}}BSD100\\ PSNR\qquad SSIM\end{tabular}} &
  {\begin{tabular}[c]{@{}c@{}}Urban100\\ PSNR\qquad SSIM\end{tabular}} &
  {\begin{tabular}[c]{@{}c@{}}Manga109\\ PSNR\qquad SSIM\end{tabular}} \\ \hline
{\begin{tabular}[c]{@{}c@{}}DANv1 \cite{8}\\ DANv2 \cite{11}\\ DCLS \cite{25}\end{tabular}} &
  {x2} &
  {\begin{tabular}[c]{@{}c@{}}37.324\qquad  0.95245\\ 37.586\qquad  0.95438\\ {\color[HTML]{3531FF} \textbf{37.607}}\qquad  {\color[HTML]{FE0000} \textbf{0.95527}}\end{tabular}} &
  {\begin{tabular}[c]{@{}c@{}}33.066\qquad  0.90426\\ 33.426\qquad  {\color[HTML]{3531FF} \textbf{0.90944}}\\ {\color[HTML]{3531FF} \textbf{33.448}}\qquad  {\color[HTML]{FE0000} \textbf{0.91044}}\end{tabular}} &
  {\begin{tabular}[c]{@{}c@{}}32.012\qquad  0.89453\\ 32.244\qquad  {\color[HTML]{FE0000} \textbf{0.89875}}\\ {\color[HTML]{3531FF} \textbf{32.261}}\qquad  {\color[HTML]{3531FF} \textbf{0.89865}}\end{tabular}} &
  {\begin{tabular}[c]{@{}c@{}}29.828\qquad  0.89261\\ 30.616\qquad  0.90568\\ {\color[HTML]{3531FF} \textbf{30.949}}\qquad  {\color[HTML]{3531FF} \textbf{0.90966}}\end{tabular}} &
  {\begin{tabular}[c]{@{}c@{}}35.972\qquad  0.96801\\ 37.005\qquad  0.97111\\ {\color[HTML]{3531FF} \textbf{37.090}}\qquad  {\color[HTML]{3531FF} \textbf{0.97172}}\end{tabular}} \\ 
{NLCUnet(Ours+)} &
  \multicolumn{1}{l}{} &
  \multicolumn{1}{l}{{\color[HTML]{FE0000} \textbf{37.759}}\qquad  {\color[HTML]{3531FF} \textbf{0.95510}}} &
  \multicolumn{1}{l}{{\color[HTML]{FE0000} \textbf{33.592}}\qquad  0.90911} &
  \multicolumn{1}{l}{{\color[HTML]{FE0000} \textbf{32.356}}\qquad  0.89855} &
  \multicolumn{1}{l}{{\color[HTML]{FE0000} \textbf{31.100}}\qquad  {\color[HTML]{FE0000} \textbf{0.91055}}} &
  \multicolumn{1}{l}{{\color[HTML]{FE0000} \textbf{37.557}}\qquad  {\color[HTML]{FE0000} \textbf{0.97272}}} \\ \hline
{\begin{tabular}[c]{@{}c@{}}DANv1 \cite{8}\\ DANv2 \cite{11}\\ DCLS \cite{25}\end{tabular}} &
  {x3} &
  {\begin{tabular}[c]{@{}c@{}}33.884\qquad 0.91965\\ 34.011\qquad 0.92086\\ {\color[HTML]{3531FF} \textbf{33.955}}\qquad {\color[HTML]{3531FF} \textbf{0.92149}}\end{tabular}} &
  {\begin{tabular}[c]{@{}c@{}}28.187\qquad 0.79365\\ 30.195\qquad 0.83102\\ {\color[HTML]{3531FF} \textbf{30.274}}\qquad {\color[HTML]{3531FF} \textbf{0.83242}}\end{tabular}} &
  {\begin{tabular}[c]{@{}c@{}}28.946\qquad 0.79365\\ 29.034\qquad 0.79661\\ {\color[HTML]{3531FF} \textbf{29.073}}\qquad {\color[HTML]{3531FF} \textbf{0.79740}} \end{tabular}} &
  {\begin{tabular}[c]{@{}c@{}}28.011\qquad 0.84343\\ 28.160\qquad 0.84776\\ {\color[HTML]{3531FF} \textbf{28.461}}\qquad {\color[HTML]{3531FF} \textbf{0.85467}}\end{tabular}} &
  {\begin{tabular}[c]{@{}c@{}}33.477\qquad 0.93965\\ 33.644\qquad 0.94155\\ {\color[HTML]{3531FF} \textbf{33.776}}\qquad {\color[HTML]{3531FF} \textbf{0.94264}} \end{tabular}} \\
{NLCUnet(Ours+)} &
  \multicolumn{1}{l}{} &
  \multicolumn{1}{l}{{\color[HTML]{FE0000} \textbf{34.273}}\qquad {\color[HTML]{FE0000} \textbf{0.92368}}} &
  \multicolumn{1}{l}{{\color[HTML]{FE0000} \textbf{30.370}}\qquad {\color[HTML]{FE0000} \textbf{0.83380}}} &
  \multicolumn{1}{l}{{\color[HTML]{FE0000} \textbf{29.150}}\qquad {\color[HTML]{FE0000} \textbf{0.79873}}} &
  \multicolumn{1}{l}{{\color[HTML]{FE0000} \textbf{28.779}}\qquad {\color[HTML]{FE0000} \textbf{0.85965}}} &
  \multicolumn{1}{l}{{\color[HTML]{FE0000} \textbf{34.092}}\qquad {\color[HTML]{FE0000} \textbf{0.94536}}} \\ \hline
{\color[HTML]{333333} {\begin{tabular}[c]{@{}c@{}}DANv1 \cite{8}\\ DANv2 \cite{11}\\ DCLS \cite{25}\end{tabular}}} &
  {x4} &
  {\begin{tabular}[c]{@{}c@{}}31.839\qquad 0.88683\\ 31.961\qquad 0.88918\\ {\color[HTML]{3531FF} \textbf{32.074}}\qquad {\color[HTML]{3531FF} \textbf{0.88937}}\end{tabular}} &
  {\begin{tabular}[c]{@{}c@{}}28.407\qquad 0.76872\\ 28.481\qquad 0.77156\\ {\color[HTML]{3531FF} \textbf{28.523}}\qquad {\color[HTML]{3531FF} \textbf{0.77271}}\end{tabular}} &
  {\begin{tabular}[c]{@{}c@{}}27.821\qquad 0.74572\\ 27.861\qquad 0.74849\\ {\color[HTML]{3531FF} \textbf{27.897}}\qquad {\color[HTML]{3531FF} \textbf{0.74885}}\end{tabular}} &
  {\begin{tabular}[c]{@{}c@{}}25.050\qquad 0.73651\\ 25.134\qquad 0.73865\\ {\color[HTML]{3531FF} \textbf{25.345}}\qquad {\color[HTML]{3531FF} \textbf{0.74657}}\end{tabular}} &
  {\begin{tabular}[c]{@{}c@{}}29.789\qquad 0.89106\\ 29.739\qquad 0.89113\\ {\color[HTML]{3531FF} \textbf{30.073}}\qquad {\color[HTML]{3531FF} \textbf{0.89686}}\end{tabular}} \\ 
{NLCUnet(Ours+)} &
  \multicolumn{1}{l}{} &
  {\color[HTML]{FE0000} \textbf{32.291\qquad 0.89317}} &
  {\color[HTML]{FE0000} \textbf{28.666\qquad 0.77510}} &
  {\color[HTML]{FE0000} \textbf{27.967\qquad 0.75180}} &
  {\color[HTML]{FE0000} \textbf{25.699\qquad 0.75971}} &
  {\color[HTML]{FE0000} \textbf{30.166\qquad 0.90048}} \\ \hline
\end{tabular}%
}
\end{table*}

\begin{figure*}[]
\centering
\includegraphics[scale=0.39]{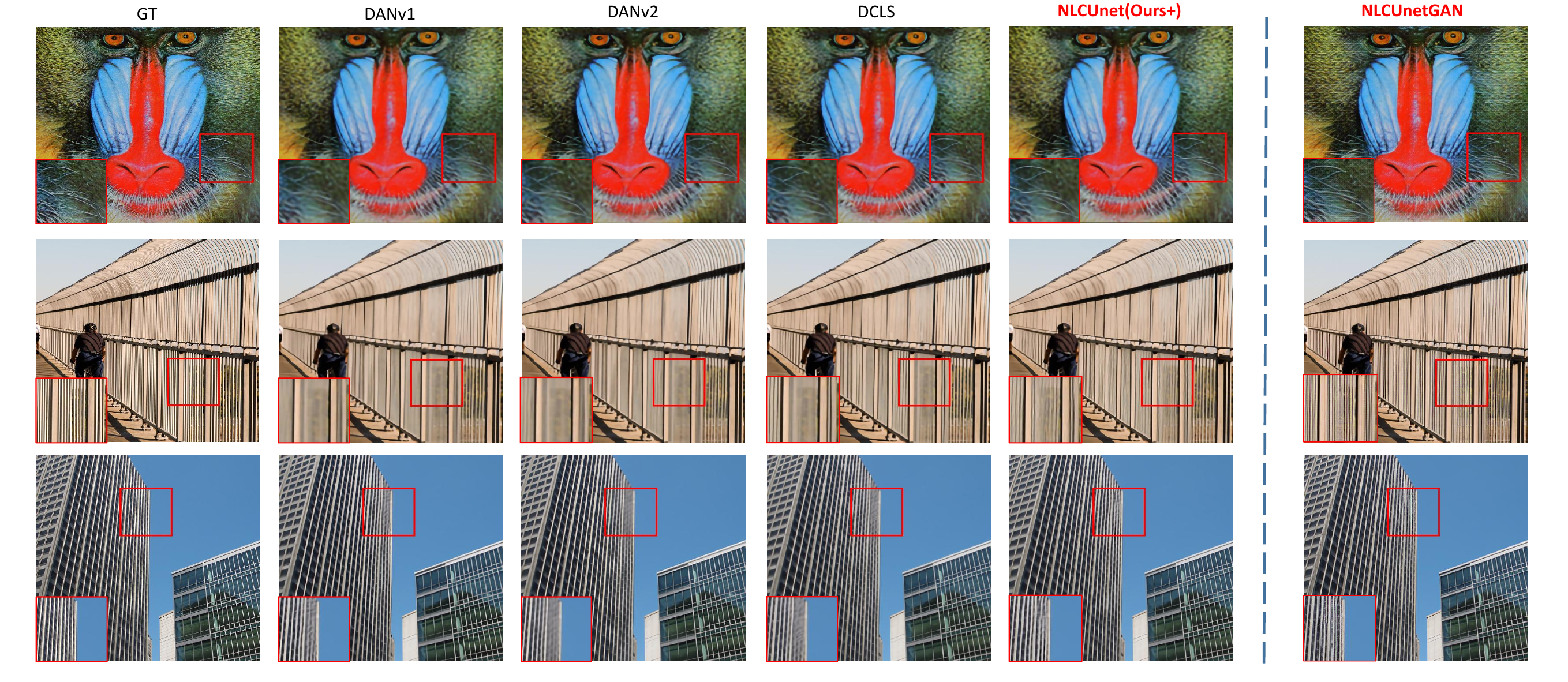}
\vspace{-0.3cm}
\caption{Visual results are from Set14 and Urban100 with scale x4 with Configuration 1. Here, the width of blur kernel is 3.2.}
\label{fig5:env}
\end{figure*}

\begin{figure*}[]
\centering
\includegraphics[scale=0.39]{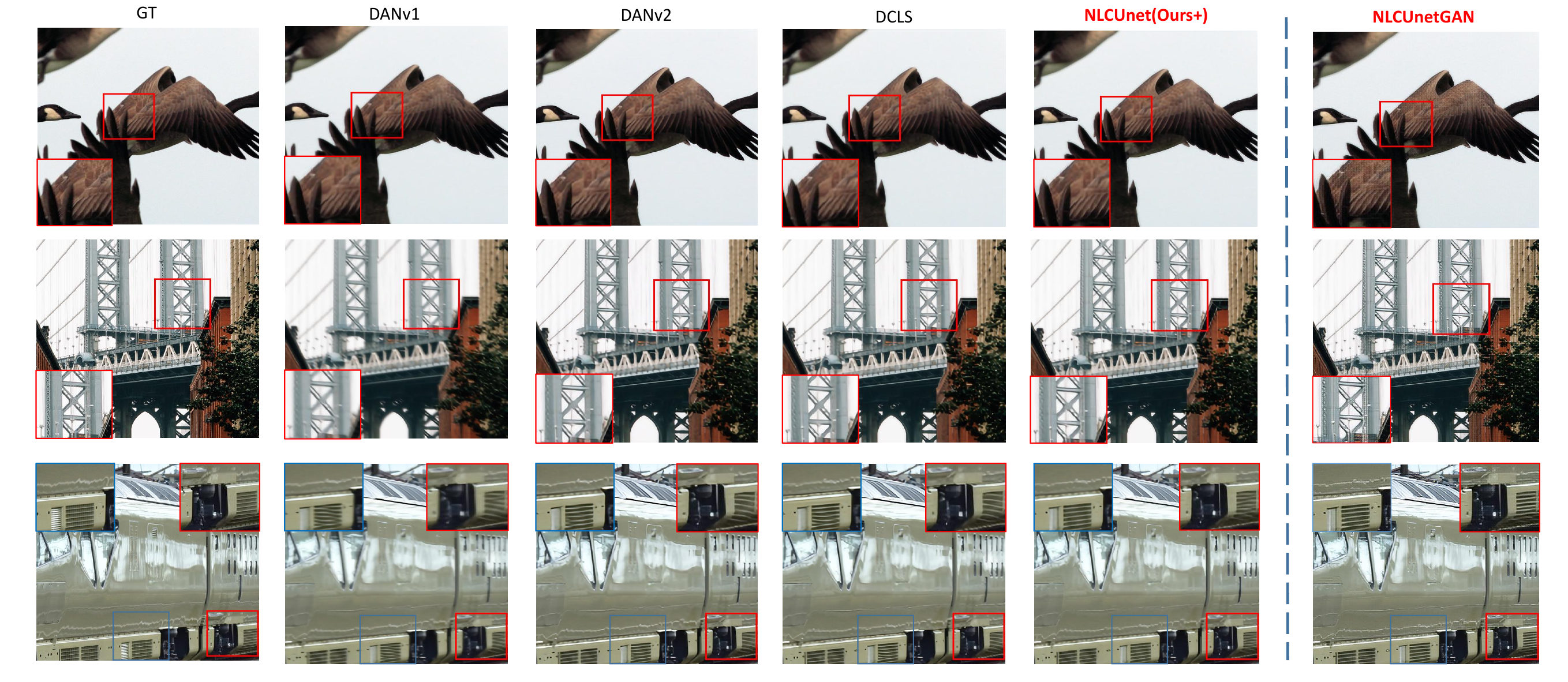}
\vspace{-0.3cm}
\caption{Visual results are from DIV2KRK with scale x4 with Configuration 2.}
\label{fig4:env}
\end{figure*}

\begin{table*}[]\tiny
\caption{Ablation experiment: the best results are highlighted in red.}
\centering
\resizebox{\textwidth}{!}{%
\begin{tabular}{@{}cccccccc@{}}
\toprule
\multicolumn{1}{l}{} &
  {DANv1} &
  \multicolumn{1}{l}{{DANv1+LN}} &
  \multicolumn{1}{l}{{DANv1(No\_Ker)+LN}} &
  \multicolumn{1}{l}{{DANv1(No\_Ker)+LN+Bic}} &
  {\begin{tabular}[c]{@{}c@{}}DANv1(No\_Ker)+LN+Bic+GDFN\end{tabular}} &
  \multicolumn{1}{c}{{\begin{tabular}[c]{@{}c@{}}Ours\\ (NLC block)\end{tabular}}} &
  \multicolumn{1}{c}{\textbf{\color[HTML]{FE0000}\begin{tabular}[c]{@{}c@{}}Ours+\\ (NLCUnet)\end{tabular}}} \\ \midrule
{PSNR (Set5)} &
  {31.83} &
  {31.85} &
  {31.84} &
  {31.89} &
  {32.00} &
  {32.14} &
  \color[HTML]{FE0000}\textbf{32.29}
   \\ \midrule
{SSIM (Set5)} &
  {0.8868} &
  {0.8869} &
  {0.8867} &
  {0.8870} &
  {0.8889} &
  {0.8907} &
  \color[HTML]{FE0000}\textbf{0.8931}
   \\ \midrule
{Gradient Explosion{?}} &
  {\Checkmark} &
  {\XSolidBrush} &
  {\XSolidBrush} &
  {\XSolidBrush} &
  {\XSolidBrush} &
  \multicolumn{1}{c}{\textbf{\XSolidBrush}} &
  \multicolumn{1}{c}{\color[HTML]{FE0000}\textbf{\XSolidBrush}} \\ \bottomrule
\end{tabular}%
}
\label{tab:my-table}
\end{table*}

\noindent{\textbf{Configuration 1.}} Following the configuration in \cite{19}, we set the size of the kernel to 21. For scale factors 2, 3, and 4, respectively, the kernel width is uniformly sampled in [0.2, 2.0], [0.2, 3.0], and [0.2, 4.0] during training. We collect HR images from the widely used benchmark datasets for quantitative evaluation, i.e., Set5 \cite{20}, Set14 \cite{21}, Urban100 \cite{22}, BSD100 \cite{23} and Manga109 \cite{24}. We uniformly select 8 kernels from the ranges [0.80, 1.60], [1.35, 2.40], and [1.8, 3.2] for scale factors 2, 3, and 4, respectively. To create synthetic test images, the HR images are first downsampled before being blurred by the chosen blur kernels.

\noindent{\textbf{Configuration 2.}} Following the option in \cite{25}, the kernel size is set to 11$\times$11 and 31$\times$31 for scale factors 2 and 4, respectively. During training, anisotropic Gaussian kernels for degradation are generated by randomly selecting a kernel width from a range (0.6, 5) and rotating from a range [$-\pi$, $\pi$]. For test scale factors 2 and 4, we cropped the DIV2KRK dataset to 256$\times$256 and 128$\times$128, respectively, and then added an anisotropic Gaussian kernel for degradation. Note that the variance of the gaussian kernel will be sampled between 0.175 and 2.5 for scale factors 2 and 4.

\indent
All scale factors have an input size of 64$\times$64 during training. The batch size is 4. our NLCUnet is trained for 1.2$\times$10$^6$ iterations, and our NLCUnetGAN is trained for 6$\times$10$^6$ iterations. We use Adam as our optimizer with $\beta_1$ = 0.9 and $\beta_2$ = 0.99. For NLCUnet, the initial learning rate is 4$\times$10$^{-4}$, and will decay by half after every 3$\times$10$^5$ iterations. For NLCUnetGAN, the initial learning rate is 1$\times$10$^{-4}$ and no decay. All models are trained on Tesla T4 GPUs.

\subsection{Experimental results and analysis}
\noindent \textbf{Analysis 1.} We test our method on test images generated by uniformly selecting 8 kernels for the first configuration. We compared three leading methods specifically made for blind SR: DANv1 \cite{8}, DANv2 \cite{11}, and DCLS \cite{25}. \textbf{Table 1} displays the PSNR and SSIM results for the Y channel of the converted YCbCr space. As shown in \textbf{Table 1}, the PSNR of the proposed NLCUnet is 0.354 dB higher than DCLS on Urban100 with scale x4, 0.318 dB higher than DCLS on Set5 and Urban100 with scale x3, and 0.467 dB higher than DCLS on Manga109 with scale x2. As shown in \textbf{Fig. 3}, the visual results are from Set14 and Urban100 with scale x4 with a blur kernel width of 3.2. As we can observe, compared to previous methods, our method significantly improves the visual results in terms of reconstructed textures, edges, and similar content.

\noindent \textbf{Analysis 2.} The second configuration, which is more commonly used but also tougher to solve compared with Configuration 1, utilizes irregular blur kernels. The PSNR and SSIM results on the Y channel are shown in \textbf{Table 3}. According to \textbf{Table 3}, the PSNR results of the proposed NLCUnet on DIV2KRK with scales x2 and x4 are greater than DCLS. We can see from the third row of \textbf{Fig. 4} that the compared methods generate tilted stripes, as highlighted in the red square. Very interestingly, our method preserves the almost same straight stripes as that of the original image. The reason is that the compared methods do not focus on the information in similar areas around the image, such as the horizontal stripes in the blue area. Instead, our method focuses on the non-local information, thus generating the same stripes as the original image.

\begin{table}[]\tiny
\centering
\caption{Quantitative comparison with three leading blind SR methods with Configuration 2. The top-1 and top-2 results are highlighted in red and blue, respectively.}
\label{tab:my-table}
\resizebox{\columnwidth}{!}{%
\begin{tabular}{ccccc}
\hline
\multirow{3}{*}{Method} & \multicolumn{4}{c}{DIVKRK}                      \\
                        & \multicolumn{2}{c}{×2} & \multicolumn{2}{c}{×4} \\
                        & PSNR      & SSIM       & PSNR      & SSIM 
                        \\
                        \hline
{DANv1 \cite{8}}                   & {27.75}   & {0.8345}   & {26.53}   & {0.7321}     \\
{DANv2 \cite{11}}                   & {28.13}   & {0.8379}   & {26.93}   & {0.7511}     \\
{DCLS \cite{25}}                   & {\color[HTML]{3531FF} \textbf{28.30}}   & {\color[HTML]{FE0000} \textbf{0.8380}}   & {\color[HTML]{3531FF} \textbf{27.02}}   & {\color[HTML]{3531FF} \textbf{0.7573}}     \\
{NLCUnet(Ours+)}          & {\color[HTML]{FE0000} \textbf{28.38}} &
  {\color[HTML]{3531FF} \textbf{0.8354}}     & {\color[HTML]{FE0000} \textbf{27.13}}     & {\color[HTML]{FE0000} \textbf{0.7595}}     \\
\hline
\end{tabular}%
}
\end{table}

\subsection{Ablation study}
The phenomenon of gradient explosion occurs in \cite{8} during training. As stated in \cite{11}, the gradient explosion was caused by the Channel Attention Layer (CALayer) in the Conditional Residual Block (CRB) of \cite{8} during training. As shown in \textbf{Table 2}, LN indicates that we add the LayerNorm operation to the model, solving the gradient explosion problem. No\_Ker indicates that we remove the blur kernel estimation module, which has almost no impact on the model. Bic indicates that we add a bicubic upsampling operation to the model, which reduces training epochs and improves performance. We used NLC as the fundamental building block of the model and Unet as the model's framework to achieve the best performance.

\indent
As shown in \textbf{Fig. 6}, the solid blue line represents that we directly apply the random cropping operation to the DF2K dataset. The solid red line means that we apply the center cropping and then random cropping to the DF2K dataset. It is obvious that the red solid line converges faster and the PSNR result is higher than that of the blue solid line. Besides, according to the experiment, our network can converge with about 6 epochs, however the compared network \cite{8} requires about 20 epochs to converge. 

\begin{figure}
    \centering
    \includegraphics[scale=0.46]{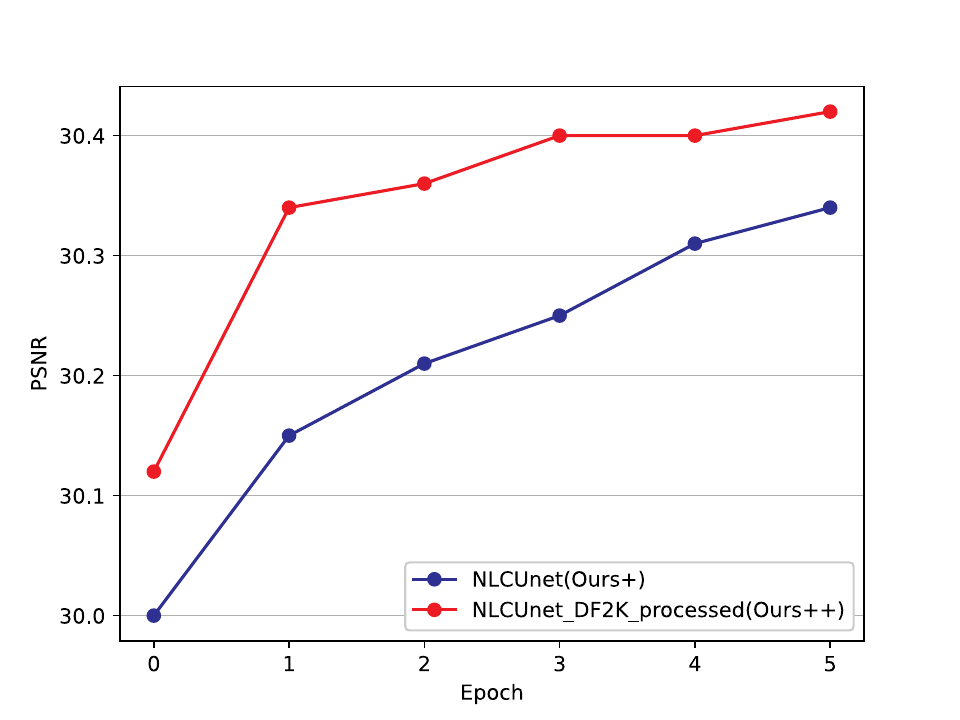}
    \caption{Average PSNR (RGB) on Set5 with scale x4}
\end{figure}

\section{Conclusion}
\label{sec:conclusion}
In this paper, we have presented a novel single-image super-resolution network by integrating GAN with non-local attention mechanism, which not only achieves good visual effects but also removes the artifacts generated solely by GAN. We design a center crop followed by a random crop to increase the semantic information of the cropped region. This greatly reduces the number of training epochs and improves performance. As a byproduct, we have discovered that adding priors to GAN yields good results in other fields, such as image de-raining and facial inpainting. In the future, we will explore the appropriate priors and apply them to the blind SR tasks.

\bibliographystyle{IEEEbib}
\bibliography{icme2023template}

\begin{thebibliography}{10}

\bibitem{1}
N.~Ahn, B.~Kang, and K.~Sohn,
\newblock ``Fast, accurate, and lightweight super-resolution with cascading
  residual network,''
\newblock in {\em European Conference on Computer Vision}, 2018, pp. 252--268.

\bibitem{2}
T.~Dai, J.~Cai, Y.~Zhang, S.~Xia, and L.~Zhang,
\newblock ``Second-order attention network for single image super-resolution,''
\newblock in {\em IEEE Conference on Computer Vision and Pattern Recognition},
  2019, pp. 11065--11074.

\bibitem{3}
J.~Kim, J.~Lee, and K.~Lee,
\newblock ``Accurate image super-resolution using very deep convolutional
  networks,''
\newblock in {\em IEEE Conference on Computer Vision and Pattern Recognition},
  2016, pp. 1646--1654.

\bibitem{4}
Y.~Xu, S.~Tseng, Y.~Tseng, H.~Kuo, and Y.~Tsai,
\newblock ``Unified dynamic convolutional network for super-resolution with
  variational degradations,''
\newblock in {\em IEEE Conference on Computer Vision and Pattern Recognition},
  2020, pp. 12496--12505.

\bibitem{5}
K.~Zhang, W.~Zuo, and L.~Zhang,
\newblock ``Learning a single convolutional super-resolution network for
  multiple degradations,''
\newblock in {\em IEEE Conference on Computer Vision and Pattern Recognition},
  2018, pp. 3262--3271.

\bibitem{6}
K.~Zhang, L.~Gool, and R.~Timofte,
\newblock ``Deep unfolding network for image super-resolution,''
\newblock in {\em IEEE Conference on Computer Vision and Pattern Recognition},
  2020, pp. 3217--3226.

\bibitem{7}
J.~Gu, H.~Lu, W.~Zuo, and C.~Dong,
\newblock ``Blind super-resolution with iterative kernel correction,''
\newblock in {\em IEEE Conference on Computer Vision and Pattern Recognition},
  2019, pp. 1604--1613.

\bibitem{8}
Y.~Huang, S.~Li, L.~Wang, T.~Tan, et~al.,
\newblock ``Unfolding the alternating optimization for blind super
  resolution,''
\newblock {\em Advances in Neural Information Processing Systems}, pp.
  5632--5643, 2020.

\bibitem{9}
E.~Agustsson and R.~Timofte,
\newblock ``Ntire 2017 challenge on single image super-resolution: Dataset and
  study,''
\newblock in {\em IEEE Conference on Computer Vision and Pattern Recognition
  Workshops}, 2017, pp. 126--135.

\bibitem{10}
R.~Timofte, E.~Agustsson, L.~Van~Gool, M.~Yang, and L.~Zhang,
\newblock ``Ntire 2017 challenge on single image super-resolution: Methods and
  results,''
\newblock in {\em IEEE Conference on Computer Vision and Pattern Recognition
  Workshops}, 2017, pp. 114--125.

\bibitem{11}
Z.~Luo, Y.~Huang, S.~Li, L.~Wang, and T.~Tan,
\newblock ``End-to-end alternating optimization for blind super resolution,''
\newblock {\em arXiv preprint arXiv:2105.06878}, 2021.

\bibitem{12}
S.~Zamir, A.~Arora, S.~Khan, M.~Hayat, F.~Khan, and M.~Yang,
\newblock ``Restormer: Efficient transformer for high-resolution image
  restoration,''
\newblock in {\em IEEE Conference on Computer Vision and Pattern Recognition},
  2022, pp. 5728--5739.

\bibitem{13}
W.~Luo, Y.~Li, R.~Urtasun, and R.~Zemel,
\newblock ``Understanding the effective receptive field in deep convolutional
  neural networks,''
\newblock {\em Advances in Neural Information Processing Systems}, pp. 1--9,
  2016.

\bibitem{14}
X.~Wang, R.~Girshick, A.~Gupta, and K.~He,
\newblock ``Non-local neural networks,''
\newblock in {\em IEEE Conference on Computer Vision and Pattern Recognition},
  2018, pp. 7794--7803.

\bibitem{15}
J.~Hu, L.~Shen, and G.~Sun,
\newblock ``Squeeze-and-excitation networks,''
\newblock in {\em IEEE Conference on Computer Vision and Pattern Recognition},
  2018, pp. 7132--7141.

\bibitem{16}
Y.~Mei, Y.~Fan, and Y.~Zhou,
\newblock ``Image super-resolution with non-local sparse attention,''
\newblock in {\em IEEE Conference on Computer Vision and Pattern Recognition},
  2021, pp. 3517--3526.

\bibitem{17}
O.~Ronneberger, P.~Fischer, and T.~Brox,
\newblock ``U-net: Convolutional networks for biomedical image segmentation,''
\newblock in {\em International Conference on Medical Image Computing and
  Computer-Assisted Intervention}, 2015, pp. 234--241.

\bibitem{18}
X.~Wang, L.~Xie, C.~Dong, and Y.~Shan,
\newblock ``Real-esrgan: Training real-world blind super-resolution with pure
  synthetic data,''
\newblock in {\em IEEE International Conference on Computer Vision}, 2021, pp.
  1905--1914.

\bibitem{19}
S.~Bell-Kligler, A.~Shocher, and M.~Irani,
\newblock ``Blind super-resolution kernel estimation using an internal-gan,''
\newblock {\em Advances in Neural Information Processing Systems}, pp. 1--10,
  2019.

\bibitem{25}
Z.~Luo, H.~Huang, L.~Yu, Y.~Li, H.~Fan, and S.~Liu,
\newblock ``Deep constrained least squares for blind image super-resolution,''
\newblock in {\em IEEE Conference on Computer Vision and Pattern Recognition},
  2022, pp. 17642--17652.

\bibitem{20}
M.~Bevilacqua, A.~Roumy, C.~Guillemot, and M.~Alberi-Morel,
\newblock ``Low-complexity single-image super-resolution based on nonnegative
  neighbor embedding,''
\newblock in {\em British Machine Vision Conference}. 2012, pp. 1--10, BMVA
  press.

\bibitem{21}
R.~Zeyde, M.~Elad, and M.~Protter,
\newblock ``On single image scale-up using sparse-representations,''
\newblock in {\em International Conference on Curves and Surfaces}, 2010, pp.
  711--730.

\bibitem{22}
J.~Huang, A.~Singh, and N.~Ahuja,
\newblock ``Single image super-resolution from transformed self-exemplars,''
\newblock in {\em IEEE Conference on Computer Vision and Pattern Recognition},
  2015, pp. 5197--5206.

\bibitem{23}
D.~Martin, C.~Fowlkes, D.~Tal, and J.~Malik,
\newblock ``A database of human segmented natural images and its application to
  evaluating segmentation algorithms and measuring ecological statistics,''
\newblock in {\em IEEE International Conference on Computer Vision}, 2001, pp.
  416--423.

\bibitem{24}
Y.~Matsui, K.~Ito, Y.~Aramaki, A.~Fujimoto, T.~Ogawa, T.~Yamasaki, and
  K.~Aizawa,
\newblock ``Sketch-based manga retrieval using manga109 dataset,''
\newblock {\em Multimedia Tools and Applications}, vol. 76, no. 20, pp.
  21811--21838, 2017.

\end{thebibliography}

\end{document}